\documentclass[11pt,a4paper]{article}
\usepackage{jheppub}

\newcommand{\rf}[1]{(\ref{#1})}
\newcommand{\beq}{\begin{equation}}
\newcommand{\eeq}{\end{equation}}
\newcommand{\be}{\begin{equation}}
\newcommand{\ee}{\end{equation}}
\newcommand{\bea}{\begin{eqnarray}}
\newcommand{\eea}{\end{eqnarray}}
\newcommand{\eq}[1]{Eq.~(\ref{#1})}
\newcommand{\non}{\nonumber \\*}
\newcommand{\ie}{{i.e.}\ }
\newcommand{\brho}{\bar\rho}
\newcommand{\blambda}{\bar\lambda}
\newcommand{\LA}{\left\langle}
\newcommand{\RA}{\right\rangle}
\newcommand{\e}{\mbox{e}}
\renewcommand{\d}{\mbox{d}}

\newcommand{\z}{\xi}

\newcommand{\om}{\omega}

\newcommand{\tr}{\mathrm{tr}\,}

\def\gtrsim{\mathrel{\mathpalette\fun >}}
\def\fun#1#2{\lower3.6pt\vbox{\baselineskip0pt\lineskip.9pt
\ialign{$\mathsurround=0pt#1\hfil##\hfil$\crcr#2\crcr\sim\crcr}}}

\def\ga{\gtrsim} 

\begin{document}


\title{The use of Pauli-Villars' regularization in string theory}

\author[a,b]
{J. Ambj\o rn}
\author[a,c]{Y. Makeenko}

\affiliation[a]{The Niels Bohr Institute, Copenhagen University,
Blegdamsvej 17,\\ DK-2100 Copenhagen, Denmark}
\affiliation[b]{IMAPP, Radboud University, Heyendaalseweg 135,\\
6525 AJ, Nijmegen, The Netherlands}
\affiliation[c]{Institute of Theoretical and Experimental Physics,
B. Cheremushkinskaya 25,\\ 117218 Moscow, Russia}

\emailAdd{ambjorn@nbi.dk} 
\emailAdd{makeenko@nbi.dk}


\abstract{The proper-time regularization of bosonic string reproduces 
the results of canonical quantization in a special scaling limit where the length
in target space has to be renormalized.
We repeat  the analysis for the Pauli-Villars regularization and demonstrate
the universality of the results.
In the mean-field approximation 
we compute the susceptibility anomalous dimension
and show it equals 1/2.
We discuss the relation with the previously known results on lattice strings. 
}


\maketitle

\section{Introduction}

Regularization plays an important role in quantum string theory. 
In the early days people mainly used the regularization by a cut-over-modes
in the mode expansion, which was intimately linked to canonical quantization.
It was already then recognized that the regularization is to be done in a covariant 
way to comply with diffeomorphism invariance. A beautiful example of why this
is important is the Brink-Nielsen computation \cite{BN73} of the energy due to
zero-point string fluctuations, which contributes both to the string tension and to
the lowest mass.
The results for the lowest (tachyonic) mass are 
reproduced also by the zeta-function regularization which has no dimensionful cut-off.

The modern path-integral string quantization~\cite{Pol81} is naturally associated
with the proper-time regularization, as the Seeley expansion of the  
matrix element of the heat kernel operator~\cite{deWitt,Gil75,DOP82,Alv83}
is used for the computation of the determinants.
It was believed for the long time that the proper-time regularization gives the same
string spectrum as canonical quantization and the zeta-function regularization.

We have recently  showed~\cite{AM15,AM17a,AM17b}  that this is indeed the case,
but only if one renormalizes length scales in target space. Using the path integral 
formulation of string theory, this is to be expected  from a field theoretical point of view 
since the target space variables $X^\mu$ are treated as ordinary
quantum fields living on the two-dimensional world sheet of the string. Generically,
one would expect such fields to receive a wave function renormalization
\beq\label{ja1}
Z^{1/2} X_R ^\mu= X^\mu,
\eeq
and this is precisely what  was needed when we performed
a mean-field calculation in bosonic string theory. From the point of view 
of perturbation theory such a mean-field calculation involves a certain 
summation to all orders of $\alpha'_0$,  and a corresponding renormalization of 
$\alpha_0'$:
\beq\label{ja2}
Z \alpha_R' = \alpha_0',\quad Z = 1 -c_1 \alpha_0' \Lambda^2 + \ldots
\eeq

Equation \rf{ja1} -- \rf{ja2} is just a standard perturbative expansion around $\alpha_0'=0$,
telling us that since the only coupling constant in the theory, $\alpha_0'$, is dimensionful,
the perturbative expansion is in $\alpha_0' \Lambda^2 
$ while target space length $X$ is naturally measured in units of $1/\sqrt{\alpha_0'}$.
However, as shown in 
\cite{AM15,AM17a,AM17b}, since the proper-time regularization provides with 
a diffeomorphism invariant cut-off $\Lambda$ and since we can calculate $Z$ 
exactly, the condition $0 \leq Z \leq 1$ forces us to  have $\alpha_0' < c_1' / \Lambda^2$. 
If we insist that $\alpha_R'$ is finite when $\Lambda \to \infty$, then $Z \to 0$
in that limit, and from the explicit expression for $Z$ we obtained  
$ Z \propto 1/\alpha_R \Lambda^2$. This implies that if $X_R$ is finite in the limit $\Lambda \to \infty$, $X$ will be of the order of $1/\Lambda$. 

The fact that the bare quantities $\alpha'_0$ and $X$ becomes singular in a limit
where the renormalized quantities are kept finite and where the cut-off is taken
to infinity should not come as a surprise from a field theoretical point of view. 
However, there are few intriguing points associated with this in 
relation to string theory.

 Firstly we have used the proper-time regularization in 
our calculation. It had build in an explicit diffeomorphism invariant 
cut-off $\Lambda$. In principle one could use a hypercubic lattice in 
target space as a regularization, the random plaquette surfaces being the string 
world sheets which appear in the path integral.  One is allowed to use such a regularization,
as well as any other regularization. It has the virtue that the Nambu-Goto action is 
simply the number of plaquettes in random surface multiplied by $a^2/\alpha_0'$.
The lattice spacing $a$ plays the role of $1/\Lambda$, but contrary to the proper-time 
regularization the cut-off refers to distances  in target space. Conceptually  this is nice 
since the world sheet is not a physical quantity which can be measured, but 
it makes it difficult to understand a relation like \rf{ja1}.

To be explicit let us consider the 
two-point function, i.e.\ the sum over all random
surfaces in the path integral formalism, 
where two points are fixed in target space and the random surfaces 
pass through these points. The standard way to define a continuum theory
from  a lattice theory is to fix a distance $|X|$ in target space, $|X| = n \cdot a$,
then let $a \to 0$, $n \to \infty$ such that $|X|$ is fixed, and then investigate 
how one should scale the dimensionless lattice coupling constants 
such that the two-point function $G(|X|)$ has a sensible limit for $a\to 0$.
For the lattice string theory the dimensionless coupling constant is 
$\mu = a^2/\alpha_0'$ and it is indeed possible to show \cite{lattice} that 
one can renormalize $\mu$ much like in \rf{ja2} such that $G(|X|)$ exists
when $a \to 0$. As shown in \cite{AM15,AM17a,AM17b} such a scaling can 
also be done in the context of our mean field theory. What is absent 
is the rescaling \rf{ja1} of $X$. Now $X$ {\it is} the physical continuum
distance and the consequence is that the two-point function $G(|X|)$
is different from the string two-point function. It is rather an ordinary 
particle two-point function \cite{lattice}.
We called this continuum world, coming from the regularized lattice 
string theory, ``Gulliver's world''. This world is in contrast to the string 
world  where a scaling \rf{ja1} of $X$ takes place. In Gulliver's world 
$X= n\cdot a$ has the extension of (infinitely) many lattice spacings when $a \to 0$,
but the scaling \rf{ja1} where $X_R$ is finite brings $X$ back to be of order $a$:
the string world becomes a ``Lilliputian world'' with the extension of a few lattice
spacings, seen from Gulliver's perspective.

Secondly, even from a purely stringy point of view there is 
something surprising about the renormalization \rf{ja1}.
Let us consider the $N$-point function for a closed string
\beq\label{ja4}
G(x_1,\ldots,x_N) = \int {\cal D} X(\omega) \; \e^{- S[X(\omega)] } \prod_{i=1}^N
\int d^2\omega_i \, \delta(X(\omega_i) -x_i) .
\eeq
For $N=2$ it is the one discussed in the previous paragraph.
While $x_1, x_2$ in ordinary quantum field theory just refer to 
spacetime coordinates  and $G(x_1, x_2)$ becomes a function of $x_1-x_2$,
in the string case they are promoted to background fields, which like
the quantum fields need a wavefunction renormalization. Thus we have the 
situation where the distances $|x_1 -x_2|$, on which $G(x_1,,x_2)$ depends,
becomes a function of the cut-off of the theory, and in fact a singular function
as our calculations in \cite{AM15,AM17a,AM17b} explicitly showed. Usually
one performs a Fourier transformation of $G(x_1,\ldots,x_N)$ in order to 
replace the $\delta$-functions in \rf{ja4} with vertex  operators, but even then 
the renormalization \rf{ja1} should be implemented  on the corresponding
momenta $P_\mu$ as
\beq\label{ja5}
P_R  = Z^{1/2} P
\eeq
in order to obtain the standard string scattering amplitude,
as described in detail in \cite{AM17b}.
The string limit is thus much more delicate than the field theoretical limit.
Again, this is of course not a surprise since the definition \rf{ja4} is in principle an off-shell
definition of the $N$-point function of string theory, and off-shell definitions are known 
to be problematic.

The  intriguing points mentioned above arise  because we have used a regularization with 
an explicit, dimensionful cut-off. In order to ensure that the results are not an 
artifact of the proper-time regularization, we have repeated the calculations using another 
regularization with an explicit, diffeomorphism invariant cut-off, namely the Pauli-Villars
regularization. In addition we calculate the string susceptibility exponent 
for our partition function. 
There are several advantages of using the Pauli-Villars regularization. First of all
and probably the most important one is that the dependence on the cut-off then becomes
explicit which allows us to use standard techniques of quantum field theory, in particular
the Schwinger-Dyson equations. Another one is that the determinants involved can be
exactly computed for certain metrics, including those for which the standard results
based on the Seeley expansion are not applicable.

We consider the Nambu-Goto string formulation, where the intrinsic world-sheet metric 
and the induced metric are treated independently by introducing a Lagrange 
multiplier. They are generically different, so only their quantum averages coincide.
We perform the path integral over target-space coordinates to obtain the effective action for
the intrinsic metric and the Lagrange multiplier, whose minimum determines their values
in the mean-field approximation which becomes exact at large number of target-space
 dimensions $d$. 
Fluctuations around the mean-field values are also governed by the effective action that
can  easily be computed to quadratic order in the fluctuations. 
Not surprisingly, the critical dimension $d=26$ plays then a crucial role.
For $d<26$ the mean field is stable under quantum fluctuations to quadratic order.
For $d>26$ it is also stable before the scaling limit is taken and in Gulliver's scaling limit,
but for $d>26$ the effective action is no longer positive definite in the Lilliputian
scaling limit. This may be associated with the presence of negative-norm states 
for $d>26$. 

In Sect.~\ref{s:mf} we formulate our setup. In Sect.~\ref{s:PV} we introduce the
Pauli-Villars regularization of  the  bosonic string and compute the effective action to
quadratic order in the  fluctuations around the mean-field values.
In Sect.~\ref{posit} we demonstrate that this action is  positive for $2\leq d <26$ 
and thus  the mean-field vacuum is stable under fluctuations.
In Sect.~\ref{s:NS} we show how an analog of the Seeley expansion of the heat kernel 
looks for the Pauli-Villars regularization. In Sect.~\ref{s:gamma} we use the standard technique of quantum field theory to calculate an 
``effective potential'' and then demonstrate an instability of the classical vacuum. We also compute the string susceptibility exponent
and obtain $\gamma_{\rm str}=1/2$ in the mean-field approximation. 
Sect.~\ref{s:dis} is devoted to a discussion of the obtained results and some speculations.
In Appendix~\ref{appA} we consider a more general 
Pauli-Villars regularization of determinants and demonstrate the universality of the results.
In Appendix~\ref{appL} we use the Gel'fand-Yaglom technique to compute 
the determinants exactly  for certain metrics and compare with the results based on the
Seeley expansion.

\section{The setup\label{s:mf}}

Let us consider a closed bosonic string in a target space with one  compactified dimension
of length $\beta$,  whose world sheet wraps once around
this compactified dimension. There is no tachyon with this setup
if $\beta$ is larger than a certain value of the order of the cut-off.
The Nambu-Goto action is given by the area of the embedded 
surface. We rewrite it, using a Lagrange multiplier $\lambda^{ab}$
and an independent intrinsic metric $\rho_{ab}$, as%
\footnote{We denote $\det  \rho=\det  \rho_{ab}$ and $\det  \lambda=\det  \lambda^{ab}$.}
\bea
S
&=&K_0 \!\int \d^2\omega\,\sqrt{\det \partial_a X \cdot \partial_bX}=
K_0 \int \d^2\omega\,\sqrt{\det  \rho} 
+\frac{K_0}2 \int \d^2\omega\, \lambda^{ab} \left( \partial_a X \cdot \partial_bX -\rho_{ab}
\right), \non
K_0&=&\frac1{2\pi \alpha'_0}.
\label{aux}
\eea
It is convenient to  choose the world-sheet coordinates  $\omega_1$ and $\omega_2$
 inside an $\omega_L\times \omega_\beta$ rectangle in the parameter space. Then
the classical solution $X^\mu_{\rm cl}$ minimizing the action \rf{aux} 
depends on $\omega$ linearly while the classical induced metric is $\omega$-independent.

Using the path-integral  quantization, 
we integrate over the quantum fluctuations of the $X$-fields by splitting
$X^\mu=X^\mu_{\rm cl} +X^\mu_{\rm q}$ and then  performing  the Gaussian path
integral over $X^\mu_{\rm q}$. 
We thus obtain the effective action,
governing the fields $\lambda^{ab}$ and $\rho_{ab}$,
\bea
S_{\rm eff}&=& K_0 \int \d^2\omega\,\sqrt{\det \rho} 
+\frac{K_0}2 \int \d^2\omega\, \lambda^{ab} \left( \partial_a X_{\rm cl } 
\cdot \partial_bX_{\rm cl } 
 -\rho_{ab} \right) 
+ \frac{d}{2}  \tr \log (-{\cal O}),  \non 
{\cal O}&  := &\frac1 {\sqrt{\det \rho} }  \partial _a \lambda^{ab} \partial_b.
\label{aux1}
\eea
The operator ${\cal O}$
 reproduces the usual two-dimensional  Laplacian for  $\lambda^{ab}= \rho^{ab}\sqrt{\det\rho}$.
Quantum observables are given by the path integral over the fields 
$\lambda^{ab}$ and $\rho_{ab}$. 
It runs over the functions $\lambda^{ab}(\omega)$ and $\rho_{ab}(\omega)$
taking on imaginary and real values, respectively.

A very important property of the quantum system with the action \rf{aux1} 
first pointed out in \cite{Pol87} is that
the field $\lambda^{ab}$ does {\em not}\/ propagate and is localized at the value
\be
\blambda^{ab}=C \rho^{ab} \sqrt{\det \rho} ,
\label{bla}
\ee
where $C$ is constant for the world-sheet parametrization we use.
We can thus rewrite the right-hand side of \eq{aux} as
\be
S=K_0 (1-C)\int \d^2\omega\,\sqrt{\det  \rho} 
+\frac{K_0 C}2 \!\int \d^2\omega\,  \sqrt{\det \rho}\,
\rho^{ab}\partial_a X \cdot \partial_b X ,
\label{auxP}
\ee
which reproduces the Polyakov string formulation~\cite{Pol81} for $C=1$.
As shown in \cite{AM15} the action \rf{auxP} is consistent only for
a certain value of $C$ which is regularization-dependent.
One has $C=1$ for the zeta-function regularization
but  $C<1$ for the proper-time regularization where
\be
C=\frac12 +\sqrt{\frac14-\frac {d\Lambda^2}{2K_0}}
\ee
as $d\to\infty$.

Instead of the proper-time regularization used in \cite{AM15,AM17a} 
we can consider 
a regularization of the Pauli-Villars type, introducing the ratio
of massless to massive determinants 
\be
{\cal R}\equiv\frac{\det(-{\cal O})\det(-{\cal O}+2M^2)}{\det(-{\cal O}+M^2)^2},
\label{newR}
\ee
when
\be
\tr\log{\cal R}= -
\int_{0}^\infty \frac{\d \tau}{\tau} \,\tr \e^{\tau{\cal O}}
\left(1-\e^{-\tau M^2}\right)^2
\label{PV22}
\ee
is convergent. Here $M\to\infty$ is the regulator mass

We have added in \rf{newR} the additional ratio of the
determinants for the masses $\sqrt{2}M$ and $M$ to cancel the
logarithmic divergence at small $\tau$, because the Seeley 
expansion
\be
\LA \omega \Big| \e^{\tau {\cal O} }
\Big| \omega \RA =
\frac1{4\pi \tau} 
\frac{\sqrt{\det{\rho}}}{\sqrt{\det{\lambda}}} +\frac {R}{24\pi}+\ldots
\label{Seeley}
\ee
starts with the term proportional to $1/\tau$.
This is specific to the two-dimensional case.
In Appendix~\ref{appA} we consider a more 
general ratio of the determinants applicable in multi-dimensional cases as well and
demonstrate universality of the results.

A nice feature of the ratio \rf{newR} is that
for some metrics, depending on only one variable,  it can be exactly
computed using the Gel'fand--Yaglom technique as is described in Appendix~\rf{appL}.
The results are compared with the ones based on the Seeley expansion \rf{Seeley}
to understand when this expansion works.

It is convenient (but not necessary) to fix the conformal gauge when
$\rho_{ab}=\rho \delta_{ab}$, so that $\sqrt{\det{\rho}}=\rho$.
Then the log of the determinant of the ghost operator
\be
{\cal O}^a_b= \left[\Delta - \frac 1{2} (\Delta \log \rho) \right] \delta^a_b
\label{Ogh}
\ee
is to be added to the effective action \rf{aux1} [or \rf{auxP}].
Equation \rf{bla} in the conformal gauge reduces to
\be
\blambda^{ab} =C\delta^{ab}
\label{cbla}
\ee
because  $\rho^{ab} \sqrt{\det \rho} =\delta^{ab} $ in the conformal gauge.

A subtlety with the computation of the determinants involved  in the conformal gauge
is now immediately seen: the fields $X^\mu$ and $\rho$ do not interact in the action~\rf{auxP}
since 
\be
S=K_0 (1-C)\int \d^2\omega\, \rho 
+\frac{K_0 C}2 \!\int \d^2\omega\,  
\delta^{ab}\partial_a X \cdot \partial_b X 
\label{auxPc}
\ee
in the conformal gauge.
But the dependence of the determinants on $\rho$ appears because the world-sheet
regularization 
\be
\varepsilon=\frac 1{\Lambda^2 \sqrt{ \det \rho} }=
\frac 1{\Lambda^2  \rho }
\ee
depends on $\rho$ owing to diffeomorphism invariance.
For smooth $\rho$ the determinants are given by the conformal anomaly~\cite{Pol81}.
An advantage of using the Pauli-Villars regularization in the conformal gauge is that 
the implicit dependence on the metric becomes explicit as we shall immediately see.

\section{Computation with the Pauli-Villars regularization\label{s:PV}}

Let us repeat the computation of the effective action of the Nambu-Goto string
to quadratic order in fluctuations
for the Pauli-Villars regularization, where $M$ in \eq{newR} plays the role of a regulator mass.
The ratio in \eq{newR} can be rewritten in the conformal gauge as 
\be
{\cal R} =
\frac{\det \left( -{\partial_a \lambda^{ab}\partial_b} \right)
\det \left(-{\partial_a \lambda^{ab}\partial_b} + 2M^2\rho\right)}
{\det \left(-{\partial_a \lambda^{ab}\partial_b} + M^2\rho\right)^2}, 
\label{PVp}
\ee
which is analogous to that for a quantum-mechanical problem in flat
space with the potential $V=M^2 \rho$. It is important that this ratio
is finite at finite $M$ and we do not have to take care of a cut-off.

For  $L\gg \beta$ we can replace one summation over modes  
by an integration and
use Plana's summation formula for the other sum over the modes. 
The finite part (the L\"uscher term)
then comes as the difference between the latter sum and the integral
as is demonstrated in Appendix~\ref{appL}, while 
the divergent as $M\to\infty$ part for
constant $\rho=\brho$ and $\lambda=\blambda $ reads
\be
\left[\log{\cal R}\right]_{\rm div}= \omega_\beta \omega_L\int \frac{\d^2 k}{(2\pi)^2}\,
\log\left[  \frac{\blambda k^2 \left(  \blambda k^2 +2M^2 \brho  \right) }
{ \left(  \blambda k^2 +M^2 \brho  \right)^2} \right] =
-M^2 \frac { \omega_\beta \omega_L \brho}{2\pi \blambda}\log 2 .
\ee
It is the same as that for the proper-time regularization with
\be
\Lambda^2 = \frac{M^2}{2\pi}\log 2 .
\label{17}
\ee


To compute the effective action to quadratic order, we expand
\be
\rho=\brho+\delta \rho, \quad \lambda^{ab}=\blambda \delta^{ab} + 
\delta \lambda^{ab} .
\ee
Every determinant can be written as the path integral
\be
\det \left(  -\partial _a \lambda^{ab} \partial_b + M^2 \rho \right)^{-d/2}=\int {\cal D} X^\mu_M
\e^{-\frac12\int \d^2 \omega\, \left(\lambda^{ab} \partial_a X_M \cdot \partial_b X_M + 
M^2 \rho X_M\cdot X_M \right)}
\ee
over the fields  $X_M^\mu(\omega)$ with normal statistics or  $Y_M^\mu(\omega)$ 
with ghost statistics.
This generates the propagator of the $X_M$ field 
\be
\LA X^\mu_M(k)X^\nu_M(-k) \RA= \frac {\delta^{\mu\nu}}{\blambda k^2 + M^2 \brho}
\ee
while the two triple vertices of the $\delta \lambda^{ab} X^\mu  X^\nu$ and 
$\delta \rho X^\mu  X^\nu$ interactions are
\bea
\LA \delta \lambda^{ab}(-p) X^\mu_M (k+p) X^\nu _M(-k) \RA_{\rm truncated}&=&
-(k+p)^a k^b\delta^{\mu\nu } ,\non
\LA\delta \rho(-p) X^\mu_M (k+p) X^\nu _M(-k) \RA_{\rm truncated}&=& - M^2 \delta^{\mu\nu } .
\eea
The latter vanishes for $M=0$  as it should owing to conformal invariance.

For the $\delta \rho \delta \rho $,   $\delta \rho \delta \lambda $ and  $\delta \lambda \delta \lambda $
terms in the effective action we find, respectively,
\bea
- \frac d2 \times\frac{\delta \rho(p) \delta \rho(-p) }2\int \frac{d^2 k}{(2\pi)^2} \frac{M^4} 
{(\blambda k^2 +M^2 \brho)[\blambda (k+p)^2 +M^2 \brho]} ,
\label{rr}\\
- \frac d2 \times \delta \lambda^{ab}(p) \delta \rho(-p) \int \frac{d^2 k}{(2\pi)^2} \frac{M^2(k+p)_a k_b} 
{(\blambda k^2 +M^2 \brho)[\blambda (k+p)^2 +M^2 \brho]} ,\label{lr} \\
-\frac d2 \times \frac {\delta \lambda^{ab}(p) \delta \lambda^{cd}(-p) }2\int \frac{d^2 k}{(2\pi)^2}
 \frac{(k+p)_a k_b(k+p)_c k_d} 
{(\blambda k^2 +M^2 \brho)[\blambda (k+p)^2 +M^2 \brho]}. \label{ll}
\eea

Let us first consider the $M^2$-term in the ratio \rf{PVp}, which is divergent as $M\to\infty$.
For $\delta \lambda^{ab}=\delta\lambda \delta^{ab}$ we find, respectively,
\bea 
\frac 12 \int \frac{d^2 k}{(2\pi)^2}\left[ \frac{4M^4} 
{(\blambda k^2 +2M^2 \brho)^2} -  \frac{2M^4} 
{(\blambda k^2 +M^2 \brho)^2} \right]&=&0, 
\label{30}\\
 \int \frac{d^2 k}{(2\pi)^2}\,k^2\left[ \frac{2M^2} 
{(\blambda k^2 +2M^2 \brho)^2} -  \frac{2M^2} 
{(\blambda k^2 +M^2 \brho)^2} \right]&=&-\frac{\Lambda^2}{\blambda^2},
\label{31} \\
\frac 12 \int \frac{d^2 k}{(2\pi)^2}\,(k^2)^2\left[ \frac1{(\blambda k^2)^2}+\frac{1} 
{(\blambda k^2 +2M^2 \brho)^2} -  \frac{2} 
{(\blambda k^2 +M^2 \brho)^2} \right]&=&\frac{\Lambda^2 \brho}{\blambda^3},
\label{32}
\eea
reproducing the expansion of
\be
-\frac d2  \Lambda^2 \int \d^2 \omega \frac {\rho(\omega) }{\lambda(\omega)}.
\ee

Let us now consider the terms ${\cal O}(p^2)$ -- the only which survive as $M\to\infty$.
It is convenient first to
integrate \rf{rr}, \rf{lr} and \rf{ll} over the angle between the vectors $p_a$ and $k_b$, 
then to
expand in $1/M$ and finally to integrate over $k^2$.
For \rf{rr} we find
\be
-\frac{d}{96\pi \brho^2} \int \d^2 \omega \, (\partial_a \delta \rho)^2
\ee
which coincides with the standard conformal anomaly 
\be
-\frac{d}{96\pi} \int \d^2 \omega \, (\partial_a \log \rho)^2
\label{35a}
\ee
to quadratic order in $\delta \rho$.
Analogously, 
we obtain from \rf{lr}
\be
-\frac{d}{24\pi \blambda \brho} \int \d^2 \omega \, (\partial_a \delta \rho) (\partial_a \delta \lambda),
\ee
which looks like quadratic order of the anomaly
\be
-\frac{d}{24\pi } \int \d^2 \omega \, (\partial_a \log \rho) (\partial_a \log \lambda),
\label{38}
\ee
Finally, for the final part of \rf{ll} we find
\be
\int \frac{\d^2 p }{(2\pi)^2}\,\delta \lambda(p)\delta\lambda(-p)
\frac{p^2 d\left[ (5-\log 8)+3 \log\frac{M^2 \brho}{p^2 \blambda}\right]}{96 \pi \blambda^2},
\label{39}
\ee
where we have assumed that $p\gg 2\pi/\omega_{\beta}$.
Notice this term is normal rather than anomalous (\ie regularization dependent).

The computation of the determinant of the ghost operator \rf{Ogh} is  similar.
It gives only the term $(\delta \rho)^2$
\be
\frac{13}{48\pi \brho^2} \int \d^2 \omega \, (\partial_a \delta \rho)^2.
\label{320}
\ee

Combining  Eqs.~\rf{35a}, \rf{38}, \rf{39} and \rf{320}, 
 we find for the effective action to quadratic order in fluctuations
\bea
\delta  S&=&-\left( K_0-\frac{d\Lambda^2 }{2\blambda^2}\right)\int \d^2\omega \, 
\delta\rho\delta \lambda
-\frac{d\Lambda^2\brho }{2\blambda^3}\int \d^2\omega \, (\delta \lambda)^2
+\frac{(26-d)}{96\pi \brho^2} \int \d^2 \omega \, (\partial_a \delta \rho)^2 \non &&
-\frac{d}{24\pi \blambda \brho} \int \d^2 \omega \, (\partial_a \delta \rho) (\partial_a \delta \lambda)
+
\int \frac{\d^2 p }{(2\pi)^2}\,\delta \lambda(p)\delta\lambda(-p)
\frac{p^2 d  \log\left(\frac{M^2 \brho}{c p^2 \blambda}\right)}{32 \pi \blambda^2},
\label{S2}
\eea
where $c$ is fixed by \eq{39}. 
This reproduces the result \cite{AM17a} for the proper-time regularization,
except for the constant $c$ in the last term which is regularization dependent.

Applying to \rf{S2} the variational derivative $-\rho(\om) \delta/\delta \rho(\om)$,
we reproduce the Seeley expansion \rf{Seeley} in the conformal gauge:
\be
\LA \omega \Big| \e^{\tau \rho^{-1} \partial _a \lambda^{ab}  \partial_b}
\Big| \omega \RA =
\frac1{4\pi \tau} \frac{\rho}{\lambda}+\frac{1}{4\pi}\left[-\frac 16 \partial_a^2 \ln \rho 
-\frac 13 \partial_a^2 \ln \lambda - \frac 14 ( \partial_a \ln \lambda)^2
\right] +{\cal O}(\tau),
\label{Seeleyc}
\ee
including the term ${\cal O}(\tau^{0})$.
 The last term on the right-hand
side of \eq{S2} does not depend on $\rho$ and thus does not contribute to
the Seeley expansion.

\section{Positivity of the effective action to quadratic order\label{posit}}

In the previous Section we have computed the effective action
assuming that $\lambda^{ab}=\lambda \delta^{ab}$.
To justify this assumption, let us consider the divergent part
of the effective action for nondiagonal $\lambda^{ab}$
\bea
S_{\rm div}&=&\int\d^2 \omega
\left[ \frac{K_0}2 \lambda^{ab} \partial_a X_{\rm cl}\cdot \partial_b X_{\rm cl}+
K_0  \rho\left(1 -\frac12  \lambda^{aa} \right) -
\frac {d \Lambda^2}2   \frac \rho{\sqrt{\det{\lambda}}}
+ \Lambda^2   \rho \right],  \non &&
\lambda^{aa}=\lambda^{11}+\lambda^{22}.
\label{cla}
\eea
It is easy to  verify this formula  for constant $\lambda^{ab}=\blambda \delta^{ab}$
and $\rho=\brho$, when~\cite{AM17a}
\begin{subequations}
\bea
\blambda&=&C\equiv \frac{1}2 +\frac{\Lambda^2}{2K_0} +
\sqrt{\frac 14\left(1 +\frac{\Lambda^2}{K_0}\right)^2 -\frac{d\Lambda^2}{2K_0}},~~
\label{newC} \\
\brho&=& \frac{L\beta}{\omega_L\omega_\beta}
\frac C{\left(2C-1-\frac {\Lambda^2}{K_0}\right) },
\label{newrho} \\
\omega_\beta&=& \frac{\omega_L}{L}\beta
\label{newbeta}
\eea
\label{mmff}
\end{subequations}
at the minimum for $\beta\gg 1/\sqrt{K_0}$.

Expanding to quadratic order 
\bea
&&\sqrt{ \det(\blambda \delta^{ab}+\delta \lambda^{ab})}=\blambda+\frac12 \delta \lambda^{aa}
-\delta \lambda_2 
+{\cal O}\left((\delta \lambda)^3\right), \non
&&\delta \lambda_2 =\frac1{8\blambda} (\delta \lambda_{11}-\delta \lambda_{22})^2+ 
\frac 1{2 \blambda}(\delta \lambda_{12})^2,
\label{3001}
\eea
we find from \rf{cla} for $\blambda=C$
\be
S^{(2)}_{\rm div} =-\frac{d\Lambda^2 \brho}{2C} \int \d^2 \omega\,\delta \lambda_2
-\left(K_0-\frac{d\Lambda^2}{2C^2}\right)\!\int \d^2 \omega \, \delta \rho \frac{ \delta \lambda^{aa}}2
-\frac{d\Lambda^2 \brho}{2C^3}\int \d^2 \omega \left( \frac{ \delta \lambda^{aa}}2\right)^2 .
\label{Sdi}
\ee

Because the path integral over $\lambda^{ab}$ goes
 parallel to imaginary axis,
\ie  $\delta \lambda^{ab}$ is pure imaginary, the exponential of the first term on the right-hand side of \eq{Sdi} (which is always {\em positive}) plays 
the role  of a functional delta-function as $\Lambda\to\infty$,
forcing $\delta \lambda^{ab}=\delta \lambda \,\delta^{ab}$. 
The last two terms on the right-hand side of \eq{Sdi} then reproduce the first
two terms in \rf{S2}.

{}From \eq{S2} for the effective action 
to the second order  in fluctuations we find the following quadratic form:
\be
\delta S_2=\int \frac{\d^2p}{(2\pi)^2} \left[
A_{\rho\rho} \frac{\delta \rho(p)\delta \rho(-p) }{\bar \rho^2} +
2A_{\rho\lambda} \frac{\delta \rho(p) \delta \lambda(-p)}{\bar  \rho \blambda} 
+A_{\lambda\lambda} \frac{\delta \lambda(p)\delta \lambda(-p) }
{\blambda^2} \right]
\label{191}
\ee
with
\be
A_{ij} = \left[ \begin{array}{cc} 
\frac {(26-d) p^2}{96\pi}& -\frac12\left(K_0-\frac{d\Lambda^2 }{2C^2}\right)\brho C
-\frac {d p^2}{48\pi}\\-\frac12\left(K_0-\frac{d\Lambda^2 }{2C^2}\right)\brho C
-\frac {d p^2}{48\pi}
&-A
\end{array}
\right],
\label{MMMM}
\ee
where
\be
A=  \frac{d \Lambda^2\bar \rho} {2C} +\frac {d p^2}{32\pi}\log (cp^2/\Lambda^2 \brho) .
\label{AAAA}
\ee
For $p^2 \ll \Lambda^2 \bar \rho$, we can drop the second term on the right-hand side of
\eq{AAAA}, so $A$ becomes constant. 
For $p^2 \ga \Lambda^2 \brho$, $A$ depends on $p^2$ but
remains positive.

Since $\delta \lambda(\omega)$ is
pure imaginary, i.e.\ $\delta \lambda (- p) = - \delta\lambda^*(p)$, we find for the determinant associated with the matrix 
 in \eq{MMMM} 
\be
D= \left[\frac12\left(K_0-\frac{d\Lambda^2}{2C^2}\right) \brho C+
\frac {d p^2}{48\pi}\right]^2+ \frac {(26-d) p^2}{96\pi} A.
\label{D}
\ee
For generic 
\be
K_0 >K_*=\left(d-1+\sqrt{d^2-2d} \right)\Lambda^2
\label{K*}
\ee
the first term in \rf{D} dominates and $D$ is positive.
We thus we a stability of the minimum \rf{mmff} with respect to quantum fluctuation

As described in detail in \cite{AM15,AM17a} the limit where the cut-off goes 
to infinity (the so-called scaling limit) is obtained by letting the bare coupling
constant $K_0$ approach $K_*$ in the following way: 
\be\label{jx1}
K_0\to K_*+ \frac{K_R^2}{2\Lambda^2 \sqrt{d^2-2d}}
\ee
for $\Lambda\to\infty$, while   $K_R$, the renormalized coupling, is fixed.
In this limit we have in addition
\be\label{jx2}
K_0-\frac{d \Lambda^2}{2C^2}\to K_R \left(1+\sqrt{1-\frac2d} \right).
\ee
This scaling is valid both in the ``Gulliver'' scaling limit and in the ``Lilliputian'' scaling 
limit. The difference between the two scaling limits is the following: 
in the Lilliputian scaling limit we scale in addition the external lengths 
$L$ and $\beta$ as $1/\Lambda$, such that $\brho$ in \rf{newrho} is finite. This implies that 
the second term on the
right-hand side of \eq{D} dominates. It is positive for $d<26$ and negative for $d>26$.
The propagator
\be
\frac{1}{\brho^2} \LA \delta \rho(p) \delta \rho(-p) \RA = \frac{48\pi} {(26-d) p^2}.
\label{32pro}
\ee
then becomes negative which may indicate a negative-norm state. 

Gulliver's scaling limit is only intended to work for the two-point function. In this 
limit $L$ and $\beta$ are not scaled like $1/\Lambda$. However, $\beta$ is 
taken to zero as  
\be\label{jx5}
\beta^2-\frac{\pi(d-2)}{3K_0 C}\propto \frac{m^2}{K_0^2}
\ee
with $m \propto \sqrt{K_R}$ being the particle mass. The interpretation 
is that the $\beta$-boundaries are contracted to points, separated by a distance $L$
in target space. In the Gulliver scaling limit we have to 
take into account the final part of the
effective action, given in the mean-field approximation by the L\"uscher term
\be
S_{\rm fin}=-\frac{\pi (d-2)}{6}\frac{\omega_L}{\omega_\beta}.
\ee
This changes Eqs.~\rf{newrho} and \rf{newbeta} to
\bea
\brho&=& \frac{L}{\omega_L\omega_\beta}
\frac{\left(\beta^2-\frac{\pi(d-2)}{6K_0 C}\right)}
{\sqrt{\beta^2-\frac{\pi(d-2)}{3K_0 C}}}
\frac C{\left(2C-1-\frac {\Lambda^2}{K_0}\right) },
\label{newrhop} \\
\omega_\beta&=& \frac{\omega_L}{L}\sqrt{\beta^2-\frac{\pi(d-2)}{3K_0 C}}.
\eea
Equations~\rf{newrho} and \rf{newbeta} are recovered for $\beta \gg 1/\sqrt{K_0}$.
 In the Gulliver  limit we have from \eq{newrhop} that
$\brho\sim \Lambda^4 $. The determinant \rf{D} is then positive for finite $p^2$,
while it changes the sign for $p^2 \sim \brho K_R^2 /\Lambda^2$ which diverges as 
$\Lambda^2$.

Both in the Gulliver and Lilliputian scaling limits $\lambda$ stays  localized, \ie $\lambda(\omega)=\blambda$. Thus only $\rho$ fluctuates.
This is similar to what is described in the book~\cite{Pol87}.

\section{Not-Seeley expansion\label{s:NS}}

The standard computation of the determinant in the proper-time regularization is
based on the Seeley expansion of the heat kernel, which emerges after applying
the variational derivative 
$-\delta/\delta \log \rho(\omega)$ to the regularized determinant.
An analogous formula for the Pauli-Villars regularization reads
\be
- \rho(\omega)\frac{\delta}{\delta \rho(\omega)} \log {\cal R}=
 \LA \omega \left| \frac {2 M^2\rho}{-\partial_a \lambda^{ab} \partial _b +M^2 \rho} \right|\omega \RA
-\LA \omega \left| \frac  {2 M^2\rho}{-\partial_a \lambda^{ab} \partial _b +2 M^2 \rho} \right|\omega \RA
 .
\label{drho}
\ee

The operator on the right-hand side of \eq{drho} is nothing but the limit of coinciding arguments of the 
matrix element 
\be
G(\omega,\omega')=\LA \omega \left|  \boldsymbol G \right|\omega' \RA
\label{52}
\ee
of the  operator
\be
\boldsymbol{ G} =  \frac {2M^2 \rho}{-\partial_a \lambda^{ab} \partial _b + M^2 \rho}
- \frac {2M^2 \rho}{-\partial_a \lambda^{ab} \partial _b + 2M^2 \rho}.
\label{53}
\ee
The role of this operator is to provide a regularization of the products of 
operators:
\be
\boldsymbol A\boldsymbol B \longrightarrow \boldsymbol A\boldsymbol G\boldsymbol B.
\ee
Using Pauli-Villars regularization $\boldsymbol{ G}$ is given by \rf{53}, while using 
the proper-time regularization it is given by the heat kernel.

For $\rho=1$ and $\lambda^{ab}=\delta^{ab}$ we have from \rf{52}, \rf{53}
a smearing of the delta function
by the difference of modified Bessel's functions
\be
G_0(\omega,0)= \frac{M^2}{\pi} \left[ K_0\left(M |\omega | \right)
 -K_0\left( \sqrt{2}M|\omega | \right)
\right]
\label{K-K}
\ee
which substitutes
\be
G_0(\omega,0) =\Lambda^2 \e^{-\pi \Lambda^2 |\omega|^2}
\ee
which appears when one uses the proper-time regularization.

The result of the Seeley expansion can be repeated for the Pauli-Villars regularization
and is given by $G(\omega,\omega)$ as shown in \eq{drho}. 
The (quadratically) divergent terms are the same provided $\Lambda^2$ and $M^2$ 
are related by 
\eq{17}. The computation of the finite term (the conformal anomaly) is  pretty much similar to
that in Sect.~\ref{s:PV}. They also coincide.

\section{The string susceptibility exponent\label{s:gamma}}

To understand the properties of the vacuum,
it is instructive to compute an ``effective potential'', like in the studies of 
symmetry breaking in quantum field theory. For this purpose we add to the action \rf{aux}
the source term
\be
S_{\rm src}=\frac {K_0}2 \int \d^2 \omega \, j^{ab} \rho_{ab}
\ee
and define the partition function $Z[j]$ in the presence of the source by path integration over
the fields.
It is clear that in the conformal gauge where
$\rho_{ab}=\rho \delta_{ab}$ and  for constant $j^{ab}=j \delta^{ab}$
this $j$  is a source for the area 
\be
A=\int \d^2 \omega\,  \rho.
\ee

Introducing the Gibbs free energy
\be
W[j]=-\frac 1{K_0 L\beta} \log Z[j]
\ee
and minimizing $W[j]$ with respect to $\brho$
for constant $j^{ab}=j \delta^{ab}$, we obtain 
\be
1+j+\frac{\Lambda^2}{K_0} -C -\frac{d\Lambda^2}{2K_0 C}=0.
\ee
Then the solution is the same as before with $C$ 
from \eq{newC} changing by
\be
C(j)=\frac{1}2 \left(1+ j +\frac{\Lambda^2}{K_0}\right)+
\sqrt{\frac 14\left(1 +j+\frac{\Lambda^2}{K_0}\right)^2 -\frac{d\Lambda^2}{2K_0}}
\label{newCj} 
\ee
while
\be
W[j]= C(j) 
\label{W=C}
\ee
in the  mean-field approximation.

From Eqs.~\rf{newCj}, \rf{W=C} we deduce 
\be
\brho(j) 
\equiv 
\frac {\partial W[j]}{\partial j}
= \frac{\partial C(j) }{\partial j}
=\frac12+
\frac { 1+j+\frac{\Lambda^2}{K_0} }{\sqrt{\left(1 +j+\frac{\Lambda^2}{K_0}\right)^2 -\frac{2d\Lambda^2}{K_0}}}
\label{rhoj}
\ee
for $\omega_L=L$ and $\omega_\beta=\beta \gg 1\sqrt{K_0}$,  reproducing \rf{newrho}  for $j=0$. 
This determines
\be
C(\brho)=\sqrt{\frac{d\Lambda^2}{2K_0}} \sqrt{\frac{\brho}{\brho-1}}
\ee
and
\be
j(\brho)=-1-\frac{\Lambda^2}{K_0}+\sqrt{\frac{d\Lambda^2}{2K_0}}
\frac{(2\brho-1)}{ \sqrt{\brho(\brho-1)}} .
\label{jjrr}
\ee

The effective potential $\Gamma(\brho)$ is defined in the standard way 
by the Legendre transformation
\be
\Gamma[\brho]\equiv W[j]
-\frac{1}{2L\beta}\int \d^2 \omega\, j^{ab} \brho_{ab} .
\label{Ga}
\ee
In the mean-field approximation  we then obtain
\be
\bar\Gamma(\brho)=
C(\brho)-j(\brho) \brho=
\left( 1+\frac{\Lambda^2}{K_0} \right) \brho -\sqrt{\frac{2d \Lambda^2}{K_0}  \brho(\brho-1)}.
\label{bGa}
\ee
Note that
\be
-\frac{\partial \bar\Gamma(\brho)}{\partial \brho}=j(\brho)
\label{10}
\ee
with $j(\brho)$ given by \eq{jjrr} as it should.

Near the classical  vacuum when $0<\brho-1\ll 1$ the potential
\rf{bGa} decreases with increasing
$\brho$ because the second term on the right-hand side has
 the negative sign. This demonstrates an 
instability of the classical vacuum. If 
$K_0>K_* $ given by \eq{K*},  
the potential \rf{bGa} increases linearly with $\brho$ for large $\brho$ and thus has a (stable) minimum at 
\be
\brho(0) =\frac 12 + \frac{1+\frac{\Lambda^2}{K_0}}
{2\sqrt{\left(1+\frac{\Lambda^2}{K_0}\right)^2-\frac {2d\Lambda^2}{K_0}}}
\label{bbrr}
\ee
which is the same as \rf{newrho} for $\beta\gg1/\sqrt{K_0}$.
Near the minimum we have
\be
\bar\Gamma(\brho)=C(0)+  \frac{K_0}{2d \Lambda^2} 
\left[ \left(1+\frac{\Lambda^2}{K_0}\right)^2-\frac {2d\Lambda^2}{K_0} \right]^{3/2}\!
\left[ \brho-\brho(0) \right]^2  +
{\cal O}\left(\left[ \brho-\brho(0) \right]^3\right).
\ee
The coefficient in front of the quadratic term is positive for $K_0>K_*$ which explicitly
demonstrates the stability of the minimum.

We can now compute a very interesting physical quantity -- the string susceptibility. 
For this purpose we define the Helmholtz free energy $ F(\brho)$
by the inverse Laplace transformation
\be
\e^{-K_0 L\beta F(\brho)}= \int \d j \,\e^{K_0 L\beta \left(j\brho-W[j]\right)},
\label{FFF}
\ee
where the integral runs parallel to the imaginary axis. The meaning of this procedure 
is a passage from grand canonical to canonical ensemble
at fixed area $A$~\cite{ADJ97}.

In the mean-field approximation we use \eq{W=C}. Then 
the integrand in \rf{FFF} has an extremum at $j(\brho)$ given by \eq{jjrr}.
Expanding about the extremum, we find
\be
j\brho- C(j) =-\bar\Gamma(\brho)+\sqrt{\frac{2K_0}{d\Lambda^2}}
\left[ \brho (\brho-1) \right]^{3/2} \left(\Delta j\right)^2.
\label{15}
\ee
The integral over $\Delta j=j-j(\brho)$ goes along the imaginary axis
and thus converges. 
For $F(\brho)$ we obtain
\be
F(\brho)= \bar\Gamma(\brho) +
\frac3{4K_0 L\beta } \log \left[ \brho (\brho-1) \right] + {\rm const.}
\label{16}
\ee

According to the definition of the string susceptibility index~\cite{ADJ97}, we expect
\be
K_0 L\beta F(\brho)={\rm regular}+ \left(  2-\gamma_{\rm str} \right) \log \frac{A}{A_{\rm min}}
\label{regular}
\ee
for ${A}\gg A_{\rm min}$. Comparing \rf{16},
this determines $\gamma_{\rm str}=1/2$.

Because the second term on the right-hand side of \eq{16} is subdominant at large $d$
(and therefore in the mean-field approximation), a question arises whether 
possible $1/d$ (or one-loop) corrections to the effective potential 
$\Gamma(\brho)$ may contribute to $\gamma_{\rm str}$.
As we shall  see momentarily, the answer is ``no''.

It is easy to compute the one-loop correction to the mean-field result \rf{bGa}.
As is shown in detail in Sect.~\ref{posit}, the only propagating field is $\delta \rho$
which results  for $d<26$ after performing the path integral over $\delta \rho$
in the standard one-loop correction to the effective action
\be
\delta S_{\rm eff}=-\frac {\Lambda^2} 2 \int \d^2 \omega \, \brho.
\ee
With the given accuracy we can identify $\brho$ in this formula with the variational
parameter to be minimized, rather than using its saddle-point value. 
Then the only effect of this additional term is to change $C(j)$, given at the saddle point by 
\eq{newC}, as
\be
C_{\rm 1loop}(j)=\frac{1}2 \left(1+ j +\frac{\Lambda^2}{2K_0}\right)+
\sqrt{\frac 14\left(1 +j+\frac{\Lambda^2}{2K_0}\right)^2 -\frac{d\Lambda^2}{2K_0}}
\label{nnewC} 
\ee
and correspondingly $\Lambda^2/K_0$ (coming from the ghosts) is substituted by 
 $\Lambda^2/2K_0$ in the above formulas. 
Notice, this is not just a simple shift $d\to (d-1)$ as one might expected. 
The critical value $K_*$, given at the saddle point by \eq{K*}, is rather changed as
\be
K_*{} _{\rm 1loop}= d-\frac 12+\sqrt{d^2-d}.
\ee
It now makes sense to consider $d>1$ like for the Polyakov string.

It is clear from this consideration that the one-loop correction contributes only to the 
regular part of $F(\brho)$ as is displayed in \eq{regular} and does not change the singular
part that gives $\gamma_{\rm str}=1/2$.

\section{Discussion\label{s:dis}}

We have applied the Pauli-Villars regularization to a relativistic string and showed
its convenience and efficiency. 
The results previously obtained with the proper-time regularization are reproduced this
way and this  demonstrates their universality. In particular, we have shown 
an instability of the classical vacuum and the stability of the mean-field vacuum
for $2\leq d<26$.

We have computed the string susceptibility exponent  in the mean field approximation and
obtained the value $\gamma_{\rm str}=1/2$. It remarkably coincides with the one
for branched polymers which can be obtained within our consideration in Gulliver's scaling
limit. The same value of $\gamma_{\rm str}=1/2$ applies also to the Lilliputian scaling
limit which corresponds to a string.

An interesting question arises as to whether the value of $\gamma_{\rm str}=1/2$ remains
valid beyond the mean-field approximation. 
We may speculate this is the case for  $2\leq d<26$
if fluctuations are described solely by the Liouville action (see \eq{Effcov}) which is quadratic
in the fields. 
But  the problem resides, as usual, in a nonlinearity of the measure for path integration 
over the Liouville field.
This issue deserves future investigation.

\subsection*{Acknowledgments}

The authors acknowledge  support by  the ERC-Advance
grant 291092, ``Exploring the Quantum Universe'' (EQU).
Y.~M.\ thanks the Theoretical Particle Physics and Cosmology group 
at the Niels Bohr Institute for the hospitality.

\appendix

\section{Universality of Pauli-Villars' regularization\label{appA}}

Keeping in mind possible applications to higher dimensions (the membranes), let us
generalize \eq{PV22} as
\be
\tr\log{\cal R}= -
\int_{0}^\infty \frac{\d \tau}{\tau} \,\tr \e^{\tau\rho^{-1} \partial_a \lambda^{ab}\partial_b}
\left(1-\e^{-\tau M^2}\right)^N
\label{PV23}
\ee
which corresponds to
\be
\tr\log{\cal R}=\sum_{n=0}^N (-1)^n C_N^n
 \log \det \left(-\rho^{-1} \partial_a \lambda^{ab}\partial_b+nM^2 \right)
\ee
with 
\be
C_N^n =\frac{N!}{n!(N-n)!}
\ee
being the binomial coefficients.
Above we worked out the case of $N=2$ but physical results should not depend on $N$.

Repeating  \rf{30} -- \rf{32}, we get
\bea 
\frac 12 \int \frac{d^2 k}{(2\pi)^2}\sum_{n=1}^N (-1)^n C_N^n
\frac{n^2 M^4} 
{(\blambda k^2 +nM^2 \brho)^2}&= &0, 
\label{30a}\\
 \int \frac{d^2 k}{(2\pi)^2}\,k^2  \sum_{n=1}^N (-1)^n C_N^n   \frac{n M^2} 
{(\blambda k^2 +nM^2 \brho)^2}&=&-\frac{\Lambda^2}{\blambda^2},
\label{31a} \\
\frac 12 \int \frac{d^2 k}{(2\pi)^2}\,(k^2)^2 \sum_{n=0}^N (-1)^n C_N^n
\frac{1} 
{(\blambda k^2 +n M^2 \brho)^2} &=&\frac{\Lambda^2 \brho}{\blambda^3},
\label{32a}
\eea
with non-universal (\ie $N$-dependent)
\be
\Lambda^2=M^2\frac{N!} {4\pi }  \int_0^\infty \d x\, 
\frac{x^2 \left[ \psi(1+N+x)-\psi(x)  \right]}{\Gamma(1+N+x)}
\ee
and
\be
\psi(x)=\frac{\d}{\d x} \log \Gamma(x).
\ee

To prove \rf{30a} we interchange the integral and the sum and rescale $k^2 \to k^2 n$ in 
each term of the sum. We then have
\be
\sum_{n=1}^N (-1)^n C_N^n n =0 .
\ee 
It is possible only in \rf{30a} but not in \rf{31a} and \rf{32a} where
the integral of each term is divergent and only the integral of the sum is convergent.

To compute the $p^2$-term, we expand
\bea 
\frac 12 \int \frac{d^2 k}{(2\pi)^2}\sum_{n=1}^N
\frac{ (-1)^n C_N^n n^2 M^4} 
{(\blambda k^2 +nM^2 \brho)(\blambda (k+p)^2 +nM^2 \brho)}&=&\frac{p^2}{48\pi\brho^2}, 
\label{30p}\\
  \int \frac{d^2 k}{(2\pi)^2}\, \sum_{n=1}^N \frac{ (-1)^n C_N^n  n M^2 k(k+p) } 
{(\blambda k^2 +nM^2 \brho)(\blambda (k+p)^2 +nM^2 \brho)}&=&-\frac{\Lambda^2}{\blambda^2}
+\frac{p^2}{12\pi\brho\blambda},
\label{31p} \\
\frac 12 \int \frac{d^2 k}{(2\pi)^2}\, \sum_{n=0}^N
\frac{ (-1)^n C_N^n  [k(k+p)]^2} 
{(\blambda k^2 +n M^2 \brho)(\blambda (k+p)^2 +nM^2 \brho)} &=&\frac{\Lambda^2 \brho}{\blambda^3}
-\frac{p^2}{16\pi\blambda^2}\log \frac{cM^2}{p^2}
\label{32p}
\eea
to order $p^2$. It becomes $p^2/n$ after $k^2 \to k^2 n$ and we find
\be
\sum_{n=1}^N (-1)^n C_N^n  =-1
\ee
both in \rf{30p} and \rf{31p}. Thus the $p^2$-terms there are universal (the conformal anomaly).
It is not the case for \rf{32p}, where the result is the log plus a non-universal constant

\section{Application of the Gel'fand--Yaglom technique\label{appL}}

The standard results for the (proper-time regularized) determinants of the two-dimensional
Laplacian with the Dirichlet boundary conditions are obtained by 
Seeley's expansion \cite{DOP82,Alv83}:
\be
\tr\log\left( -\Delta \right)\Big|_{\rm div}
=-\frac 1{4\pi} \left\{
\Lambda^2\int_D-\sqrt{\pi} \Lambda \int_{\partial D}+\frac 13\log{\Lambda^2}
\left[\int_D \frac R2 +\int_{\partial D} k
\right] \right\}
\label{Gilkeyd}
\ee
for the divergent part and
\be
\tr\log\left( -\Delta\right)\Big|_{\rm fin}=-\frac 1{24\pi} \left[
\int_D \frac 12 R\phi+\int_{\partial D} k\phi
\right]-\frac1{4\pi} \int_{\partial D} k  
\label{Gilkey}
\ee
for the finite part in the conformal gauge $\rho_{ab}=\e^{\phi} \delta_{ab}$. Here
\be
k=-\frac 12 n^a \partial _a  \phi
\label{k}
\ee
is the geodesic curvature and $n^a$ is
the inward normal unit vector. 


The action describing dynamics of the Liouville field $\phi$  
in the Polyakov string formulation emerges from path
integration over $X^\mu$ (and the ghosts)
due to ultraviolet divergences regularized
by a cut-off. For smooth $\phi$ its finite bulk part is given by the conformal anomaly 
\beq
S_{\rm L}=\frac{d-26}{96\pi} \int   
R\Delta^{-1} R =
\frac{26-d}{96\pi} \int d^2 \omega \,(\partial_a \phi)^2.
\label{Effcov}
\eeq
This  formula is applicable for smooth metrics with the curvature
\be
R\ll \Lambda^2 ,
\label{ineq}
\ee 
when the determinants result in the conformal anomaly.
However, in the path integral over $\phi$ we integrate, in particular,
over $\phi$'s for which the inequality \rf{ineq} is not satisfied.

A simplest example of these are discontinuous metrics when $R$ is infinite
at the discontinuities, so that
\rf{ineq} is not satisfied. For $d>26$ they will dominate the path integral with the
action~\rf{Effcov} because of the negative sign. This is a disastrous feature of
the Liouville action for $d>26$.
It is to be compared with the role plays by discontinuous 
trajectories in the Brownian motion, where they are suppressed because of 
the positive sign of the action.
Thus a question arises as to whether we can indeed approximate the exact
effective action by the conformal anomaly for this kind of metrics.

Below in this Appendix we shall exactly compute  the determinants 
for particular metrics: both for the case of a smooth $\phi$ where \eq{Effcov}
works (Subsect.~\ref{ss:co}) and a discontinuous  $\phi$ where \eq{Effcov} does not work
(Subsect.~\ref{ss:dico}), using the  Gel'fand-Yaglom technique reviewed in 
 in Subsect.~\ref{ss:G-Y}. We shall compare the results with 
Eqs.~\rf{Gilkeyd}, \rf{Gilkey} and find an agreement when $\phi$ is smooth.
If $\phi$ is discontinuous, we shall see an essential difference between an exact 
result for the determinant and \eq{Gilkey}.

\subsection{The Gel'fand-Yaglom technique\label{ss:G-Y}}

The ratio in \eq{PVp} 
\be
{\cal R} =
\frac{\det \left( -\partial^2 \right)
\det \left(-\partial^2 + 2M^2\e^{\phi}\right)}
{\det \left(-\partial^2 + M^2\e^{\phi}\right)^2}, 
\label{PVpp}
\ee
is analogous to that for a quantum-mechanical problem in flat
space with the potential $V=M^2\e^\phi$. It is important that this ratio
is finite and we do not have to take care of a cut-off.

The ratio of the determinants in \eq{PVpp} can be computed for
the Dirichlet boundary conditions in some cases 
by the Gel'fand--Yaglom technique.
Let us consider the coordinates on a strip: $x\in[x_0,x_1]$,
$\theta \in [0,2\pi]$, and choose 
\be
\phi(x,\theta)=\varphi(x)
\label{choose}
\ee 
that depends only on $x$. Expanding in modes $\e^{i n \theta}$,
we can rewrite the ratio of 2d determinants in \eq{PVpp} as a product of
the ratios of 1d determinants
\be
{\cal R}^{(1)}\equiv
\frac{\det \left( -\partial^2 \right)} {\det \left(-\partial^2 + M^2\e^{\varphi}\right)}
= \prod_n
\frac{\det \left( -\partial^2_x  +n^2\right)}
{\det \left(-\partial^2_x+n^2 + M^2\e^{\varphi}\right)}. 
\label{detn}
\ee

The ratio of the 1d determinants on the right-hand side of \eq{detn}
is then given by the ratio 
\be
\frac{\det \left(-\partial^2_x+n^2 + M^2\e^{\varphi}\right)}
{\det \left( -\partial^2_x  +n^2\right)}=
\frac{\Psi_{n}(x_1)}{\Psi_{n}^{\rm free}(x_1)} 
\label{detnp}
\ee
of the properly normalized solutions ($ \Psi_n(x_0)=0, \Psi'_n(x_0)=1$) to the 
Schr\"odinger equations with zero eigenvalues, while the 
solution in the free case reads
\be
\Psi_{n}^{\rm free}(x)= \frac {\sinh[n(x-x_0)]}{n}.
\ee

\subsection{Continuous metric\label{ss:co}}

We shall elaborate on the case, when
\be
\e^\varphi =x ~~\qquad(0<x_0\leq x \leq x_1 ) . \label{c3}
\ee
The metric \rf{c3} is increasing with $x$, so 
we might expect no deviations from the standard results unless
$x_0$ is very small, \ie when  $M^2 x_0$ is no longer large.
For the metric \rf{c3} the curvature $R=x^{-3}$ becomes large as $x_0\to 0$
violating \rf{ineq}.
We shall 
fix the scaling factor of the metric by requiring the area to be equal to $2\pi$. 
This implies  $x_1=\sqrt{2}$ for \rf{c3}. 

The associated solution reads
\be
\Psi_{n}(x)=\frac{2}{3M^{2/3}}\sqrt{\z}\sqrt{\z_0}\left[
I_{1/3}\left(\frac 23 \z^{3/2}\right)K_{1/3}\left(\frac 23 \z_0^{3/2}\right)
-K_{1/3}\left(\frac 23 \z^{3/2}\right)I_{1/3}\left(\frac 23 \z_0^{3/2}\right)
\right]. 
\label{s3} 
\ee
We have used here a representation of
the Airy functions through the modified Bessel functions and denoted
\be
\z=\frac{n^2+M^2 x}{M^{4/3}},\qquad \z_0=\frac{n^2+M^2 x_0}{M^{4/3}}.
\label{zvsx}
\ee

The usual quadratic divergence of 2d determinants cancels in the
ratio on the right-hand side of \eq{detn}, while the logarithmic 
divergence is of the form
\be
{\cal R}^{(1)}\Big|_{\rm div}=
\exp\left[\frac 1{2\pi}
\left(\sum_{n=1}^\infty \frac 1n -\gamma_{\rm E} \right)
\int \d^2 z\, M^2 \e^{\phi}
\right],
\label{logdiv}
\ee 
where the Euler constant $\gamma_{\rm E}$ emerges because of the
difference between the sum and the integral.
This logarithmic divergence will cancel out in the ratio \rf{PVpp},
but is present for the ratio \rf{detn} as is already mentioned.

The coefficient of the logarithmic divergence in \eq{Gilkeyd} involves 
the Euler character
\be
\chi=\frac1{2\pi}\left( \frac12 \int_D R+ \int_{\partial D} k \right). 
\ee
For $\phi=\varphi(x)$ we have
\be
\frac1{4\pi} \int_D R =-\frac12 \int_{x_0}^{x_1} \d x \, \partial_x^2 \varphi(x)
=- \frac12 \partial_x \varphi(x) \Big |_{x_0}^{x_1} 
=- \frac1{2\pi} \int_{\partial D} k  
\ee
so that $\chi=0$. This means that we deal with an upper half plane for a
periodical real axis, which is then conformally mapped onto a strip.
Analogously, integrating by parts, we have
\be
\frac1{2\pi}\left( \frac12 \int_D R \varphi+\int_{\partial D} k \varphi\right)
=\frac12\int_{x_0}^{x_1} \d x \left(\partial_x \varphi\right)^2.
\ee

Subtracting the logarithmic divergence \rf{logdiv},
we arrive for the ratio of the determinants at the products
\bea
{\cal R}^{(1)}\Big|_{\rm fin}= \e^{\gamma_{\rm E} \int^{x_1}_{x_0} d x\, M^2 e^{\varphi(x)}}
\prod_n \frac{n}{\sinh[n(x_1-x_0)]}\,\e^{-\int^{x_1}_{x_0} d x \, 
M^2 e^{\varphi(x)}/2n}\Psi_n(x_1)
\eea
which are convergent. For the solution \rf{s3} this can be
explicitly verified by substituting its proper asymptote as
$n\to \infty$. 

For the solution \rf{s3} with $M\gg1$ we obtain
\bea
\lefteqn{{\cal R}=\e^{\gamma_{\rm E}  M^2(x_1^2-x_0^2)/2} }\non &&\times
\prod_n \frac{n}{\sinh[n( x_1-x_0)]}\e^{ -M^2(x_1^2-x_0^2)/4n}  
\frac{2}{3M^{2/3}}\sqrt{\z_1}\sqrt{\z_0}\,
I_{1/3}\left(\frac 23 \z_1^{3/2}\right)K_{1/3}\left(\frac 23 \z_0^{3/2}\right).
\non &&
\label{r3}
\eea
This product can hopefully be evaluated using Plana's summation formula
\be
\frac 12 f(0)+\sum_{n=1}^\infty f(n)=\int_0^\infty \d \om\,f(\om)+
\i\int_0^\infty \d t\, \frac {f(\i t)-f(-\i t)}
{\e^{2\pi t}-1},
\label{Plan}
\ee
which holds when $f(z)$ is analytic for ${\rm Re}\; z\geq 0$, in particular,
at the imaginary axis. 

For the solution \rf{s3} the limits $n\to\infty$ and $M\to\infty$
commute  if  $x_0\gg M^{-2/3}$, which is {\em precisely}\/
when \rf{ineq} is satisfied, and we can substitute the (modified) Bessel
functions by their asymptotic expansions
\bea
I_{1/3}\left(\frac 23 \z^{3/2}\right)&=&
\sqrt{\frac{3}{4 \pi \z^{3/2} } }\e^{2 \z^{3/2}/3} 
\left(1+\frac{5}{48 \z^{3/2}}+{\cal O}\left(\z^{-3}\right)\right),\non
K_{1/3}\left(\frac 23 \z^{3/2}\right)&=&
\sqrt{\frac{3 \pi}{4 \z^{3/2} } }\e^{-2 \z^{3/2}/3} 
\left(1-\frac{5}{48 \z^{3/2}}+{\cal O}\left(\z^{-3}\right)\right).
\label{Expa}
\eea  
The next terms of the expansions will not effect the $M\to \infty$ limit to be
taken after the computation of the product over $n$ by using
\eq{Plan}. 
Analogously, the second integral on the right-hand side of
\eq{Plan} is exponentially suppressed as $M\to\infty$.
Inserting the expansion \rf{Expa} into the first integral on the right-hand 
side of \eq{Plan}, we obtain as $M\to \infty$  for the final part of the 
product
\be
\log{\cal R}^{(1)}\Big|_{{\rm fin}} = \frac 5{24}
\left(\frac{1}{x_0}- \frac{1}{x_1}\right). 
\label{someRes}
\ee
The same formula obviously holds for the ratio \rf{PVpp}.

This is to be compared with the value of the Liouville action
\be
-\frac1{48\pi} \int \d^2 z\, \partial _a \phi \partial _a \phi =
-\frac1{24} \int_{x_0}^{x_1} \d x\, \partial _x \varphi\partial _x \varphi 
=-\frac1{24} \left(\frac{1}{x_0}- \frac{1}{x_1}\right)
\label{l3}
\ee
for $\phi=\log x$. The obtained structure is similar, while the
difference of the coefficients is due to the
boundary term that reads [see Eq.~(4.42)
of \cite{Alv83} with $\sigma=\phi/2$]
\be
\frac1{8\pi} \int \d s\,\e^{\phi/2} n^a \partial_a \phi =
\frac1{8\pi} \int_0^{2\pi} \d \theta
\left[ \partial_x \varphi(x)\Big|_{x=x_0} -
\partial_x \varphi(x)\Big|_{x=x_1} \right] =
\frac1{4} \left(\frac{1}{x_0}- \frac{1}{x_1}\right).
\label{b3}
\ee
The sum of \rf{l3} and \rf{b3} indeed coincides with \rf{someRes}, so is that it agrees with
the standard result \rf{Gilkeyd} and \rf{Gilkey} when \rf{ineq} is satisfied.

\subsection{Discontinuous metric\label{ss:dico}}

Let us consider the case when $\phi$ is constant along
$\omega_2$  and has a discontinuity from $\varphi_1=0$ to $\varphi_2>0$
at a certain value of $\omega_1$. Since 
\be
\int d^2 \omega (\partial_a \phi)^2 \propto \frac \beta \delta
\left( {\varphi_2}-  {\varphi_1}\right)^2
\label{diver}
\ee
is divergent when we vanish smearing $\delta$ of the discontinuity,
one might think this leads to an instability for $d>26$.
But $\det (-\Delta)$ for such a discontinuous metric is larger than
one for constant $\varphi=\varphi_2$, because all eigenvalues are larger.
It cannot thus be zero as the Liouville action says. 

Using the Gel'fand-Yaglom technique, we can explicitly compute the
ratio \rf{PVpp} for such a metric, which is constant along the periodic
coordinate and discontinuous along another one. Let $\omega_1\equiv x$
ranges from $0$ to $L$ and the metric has a step from 
$\e^{\varphi_1}$ to $\e^{\varphi_2}$ at a certain intermediate value
 $x=x_i$. The proper solution reads
\be
\Psi_n(x)=\left\{
\begin{array}{ll}
\frac1{m_1}\sinh m_1 x &  0\leq x\leq x_i \\
\frac{1}{m_2}\cosh (m_1 x_i)\sinh [m_2(x-x_i)] & \\
~~+\frac{1}{m_1}\sinh (m_1 x_i)\cosh [m_2(x-x_i)]
 & x_i\leq x\leq L \\
\end{array}
\right.
\label{ppsi}
\ee
where we set $\beta=2\pi$ and
\be
m_1=\sqrt{n^2+\e^{\varphi_1} M^2},\qquad m_2=\sqrt{n^2+\e^{\varphi_2} M^2}.
\ee
For $x_i=L/2$ this gives
\be
\Psi_n(L)
=\frac 12 \left( \frac{1}{m_1}+   \frac{1}{m_2}\right)
\sinh \frac{L(m_1+m_2)}2+\frac 12 \left( \frac{1}{m_1}-   \frac{1}{m_2}\right)
\sinh \frac{L(m_1-m_2)}2.
\label{PhiL}
\ee

For asymptotically large $L$ and $x_i\sim L$ we get from \eq{ppsi} for the bulk term
\be
\ln \frac{\Psi_n(L)}{\Psi_n^{\rm free}(L)} = (m_2-n) L +(m_1-m_2) x_i
\ee
and
\be
\ln {\cal R} = \sum_{n=-\infty}^{+\infty} \left[(n-m_2) L +(m_2-m_1) x_i \right].
\ee

To compute the sum, we use Plana's summation formula \rf{Plan},
where the second term on the right-hand side
describes the difference between the sum and the integral.
In our case this term gives the standard result
\be
-2L\int_0^\infty \d t \frac{t}{\e^{2\pi t}-1}=-\frac L{12}
\ee
modulo exponentially small terms $\sim L\e^{-2\pi M}$,
which arise from the domain $t>M$.

The computation of the $M$-dependent part is based on the integral
\be
\int_0^\infty \d x \left( x-2 \sqrt{x^2+A}+\sqrt{x^2+2A}\right) =- \frac A2 \ln 2
\ee
For the bulk part of the ratio \rf{PVpp} it gives
\be
\tr \ln(-\Delta)|_{\rm reg}=-\frac{\beta M^2}{4\pi}\ln2 \left[\e^{\varphi_1} x_i +\e^{\varphi_2} (L-x_i)
\right] -\frac{\pi L}{6\beta}
\ee
The singular at $M\to\infty$ part is of the type $M^2 \int \d^2\omega \,\e^\phi$ as it should.

The boundary term, which may potentially diverge like in \eq{diver}, comes from the pre-exponential in \eq{ppsi}:
\bea
\lefteqn{\sum_n \log \left[ \frac n2 \left( \frac1{m_1}+  \frac1{m_2}\right) \right] }\non &&=
\sum_n \log \left[ \frac n2 \left( \frac1{\sqrt{n^2+\beta^2 M^2\e^{2\varphi_1}/4\pi^2}}+ 
 \frac1{{\sqrt{n^2+\beta^2 M^2\e^{2\varphi_2}/4\pi^2}}}\right) \right] .
\label{34}
\eea
For the part which becomes divergent as $M\to\infty$, we can
replace the sum by an integral to obtain the complete elliptic integral of the second kind:
\bea
\frac{ \beta M}\pi \int_0^\infty \d x \,\log \left[ \frac 12 \left( \frac1{\sqrt{1+\e^{2\varphi_1}/x^2}}+ 
 \frac1{{\sqrt{1+\e^{2\varphi_2}/x^2}}}\right) \right] \non =
\frac{ \beta M}\pi   \e^{\varphi_2}
\left[ 2 E\left(\sqrt{1-\e^{2(\varphi_1-\varphi_2)}}\right)   -\frac \pi 2 
\left( 1+\e^{\varphi_1-\varphi_2} \right)\right].
\label{35}
\eea
When the discontinuity vanishes (\ie $\varphi_2=\varphi_1$), we find
\be
(\ref{35})\to \frac{ \beta M}2   \e^{\varphi_1}
\ee
which determines the boundary term in $\tr \log{\cal R}^{(2)}$ to be
\be
\left( 2-\sqrt{2} \right)\frac{ \beta M}2   \e^{\varphi_1}.
\ee
The sign is positive as it should be for the Derichlet boundary condition.

Equation~\rf{35} shows how this term is modified for $\varphi_2>\varphi_1$, but
it definitely remains finite as was anticipated by the inequality below \eq{diver}.
Therefore, the reason why the Liouville action was divergent for the discontinuous metric
is that the limit of $\Lambda\to\infty$ is {\em not}\/ interchangeable with the limit of the smearing
parameter $\delta\to0$. In \eq{diver} the limit $\Lambda\to\infty$ was taken first, while in \eq{35}
the limit $\delta\to0$ was taken first.
Thus there is no divergence for the above discontinuous metric in a regularized theory.

\end{document}